\documentclass[letterpaper, 10 pt, conference]{ieeeconf}  
\IEEEoverridecommandlockouts                              
\usepackage[
draft,
pdfpagelabels=false,
pdfpagelabels=false,
hyperindex=false,
pageanchor=false
]{hyperref}
\usepackage{lineno}
\usepackage{amsmath}
\newtheorem{theorem}{Theorem}
\newtheorem{remark}{Remark}
\usepackage{xcolor}
\usepackage{booktabs}
\usepackage{pgfplots}
\usepackage{cite}
\usepackage{enumerate}
\usepackage{multirow}

\newcommand{\refpos}{\phi}
\newcommand{\pedal}{\rm pad}
\newcommand{\lung}{\rm L}
\newcommand{\ambu}{\rm bag} 
\newcommand{\dV}{{\rm d}V}
\newcommand{\reference}{\rm ref} 
\newcommand{\target}{\rm sp}

\title{
\Huge %
Volume Control of Low-Cost Ventilator with \\ Automatic Set-Point Adaptation
}

\author{Lukas Hewing\textsuperscript{*}, Marcel Menner\textsuperscript{*}, Nikolaos Tachatos, Marianne Schmid Daners, \\ Cosima du Pasquier, Thomas S. Lumpe, Kristina Shea, Andrea Carron, and Melanie N. Zeilinger
\thanks{This work was supported by ETH Zurich, financially and through lab access.}
\thanks{\textsuperscript{*} Co-first authorship.}%
\thanks{L. Hewing, M.~Menner, A. Carron, and M.~N.~Zeilinger are with the Institute for Dynamic Systems and Control, ETH Zurich, 8092 Zurich, Switzerland (e-mail: lhewing@ethz.ch; mmenner@ethz.ch; carrona@ethz.ch; mzeilinger@ethz.ch).}%
\thanks{N. Tachatos and M. Schmid Daners are with the Product Development Group Zurich, ETH Zurich, 8092 Zurich, Switzerland (e-mail: tachaton@ethz.ch; marischm@ethz.ch).}%
\thanks{C. du Pasquier, T. S. Lumpe, and K. Shea are with the Engineering Design and Computing Laboratory, ETH Zurich, 8092 Zurich, Switzerland (e-mail: cosimad@ethz.ch; tlumpe@ethz.ch; kshea@ethz.ch).}%
}

\begin{document}

\maketitle
\thispagestyle{empty}
\pagestyle{empty}

\begin{abstract}

This paper considers the control design for a low-cost ventilator that is based on a manual resuscitator bag (also known as AmbuBag) to pump air into the lungs of a patient who is physically unable to breathe. First, it experimentally shows that for accurately tracking tidal volumes, the controller needs to be adapted to the individual patient and the different configurations, e.g., hardware or operation modes. Second, it proposes a set-point adaptation algorithm that uses sensor measurements of a flow meter to automatically adapt the controller to the setup at hand. Third, it experimentally shows that such an adaptive solution improves the performance of the ventilator for various setups. One objective of this paper is to increase awareness of the need for feedback control using sensor measurements in low-cost ventilator solutions in order to automatically adapt to the specific scenario.

\end{abstract}


\vspace{0.1cm}

\section{Introduction}

Coronavirus disease 2019 (COVID-19) is an ongoing pandemic, which is caused by severe acute respiratory syndrome coronavirus 2 (SARS-CoV-2).
The viral spread of the disease has dramatically increased the demand for medical devices needed in hospitals to treat patients.
Mechanical ventilators are one example of medical devices that are currently in short supply and that are critical for the treatment of patients suffering from COVID-19, cf. \cite{pearce2020review}.
To meet the peak ventilator demands, a series of low-cost and modular solutions have been proposed, e.g., in \cite{li2020utah, galbiati2020mechanical}.
In order to provide guidelines on their design, the Medicines and Healthcare products Regulatory Agency \cite{UKguidelines} has introduced requirements that a low-cost ventilator solution ought to satisfy.
These guidelines include hardware requirements, operation modes, and settings, as well as alarms for the malfunctioning of the ventilator.

A popular class of low-cost mechanical ventilators is a design that relies on squeezing a manual resuscitator bag (AmbuBag) with paddles to pump air into the lungs of a patient.
Such a ventilator solution is mechanically flexible enough to satisfy the operational requirements but depends on an adequate control design to actuate the paddles. 
This paper addresses the controller design for a bag-based ventilator building on \cite{MITevent} design where we particularly focus on satisfying the following four requirements:
\begin{enumerate}[i.]
\item 
Positive end-expiratory pressure (PEEP).
The ventilator must provide a range of 5--20~mBar during expiration, adjustable in 5~mBar increments. 
\item
Inspiratory:Expiratory ratio (I:E). The ventilator must provide a ratio of 1:2, and additionally could  provide a range of 1:1--1:3.
\item
Breaths per minute (BPM).
The ventilator must provide a respiratory rate of 10--30 BPM, adjusted in 2~BPM increments.
\item 
Tidal volume. 
The ventilator must provide a tidal volume of 400~mL.  
Additionally, it should provide tidal volumes of 350 and 450~mL.
\end{enumerate}
For an extensive list of all requirements, the reader is referred to \cite{UKguidelines}.

We investigate the influence of different patients and different settings on the operation of a bag-based low-cost ventilator.
In particular, we show experimentally that feedforward control of the paddles may not be sufficient to achieve accurate tidal volume tracking (requirement iv.) for different operational settings (e.g., along requirements i.-iii.) or patient characteristics, due to varying compression and leakage losses.
Instead, we propose to utilize feedback from a flow meter to adapt the controller to the individual patient and the specific operational settings, as well as small deviations in the mechanical set-up (bag placement, tubing, etc.).

The experiments were conducted using a mechanical lung (TestChest V2, Organis GmbH, Landquart, Switzerland) 
that is capable of simulating different patients and healthy persons. 
With this study, we want to emphasize the importance of a flow meter and an adaptation of the controller, regarding patient safety and alarm signals.
Related ventilator designs do not mention the implementation of a flow meter or an adaptive scheme to cope with a patient's pathophysiology, cf.~\cite{ApolloBVM,Coventor,galbiati2020mechanical,MITlowcost2010,li2020utah,MITevent,SpiroWave}.

\section{Methods}

\subsection{The Breathe Ventilator}
\begin{figure}[t!]
      \centering
\begin{tikzpicture}
     \node{\includegraphics[
     width=7cm]{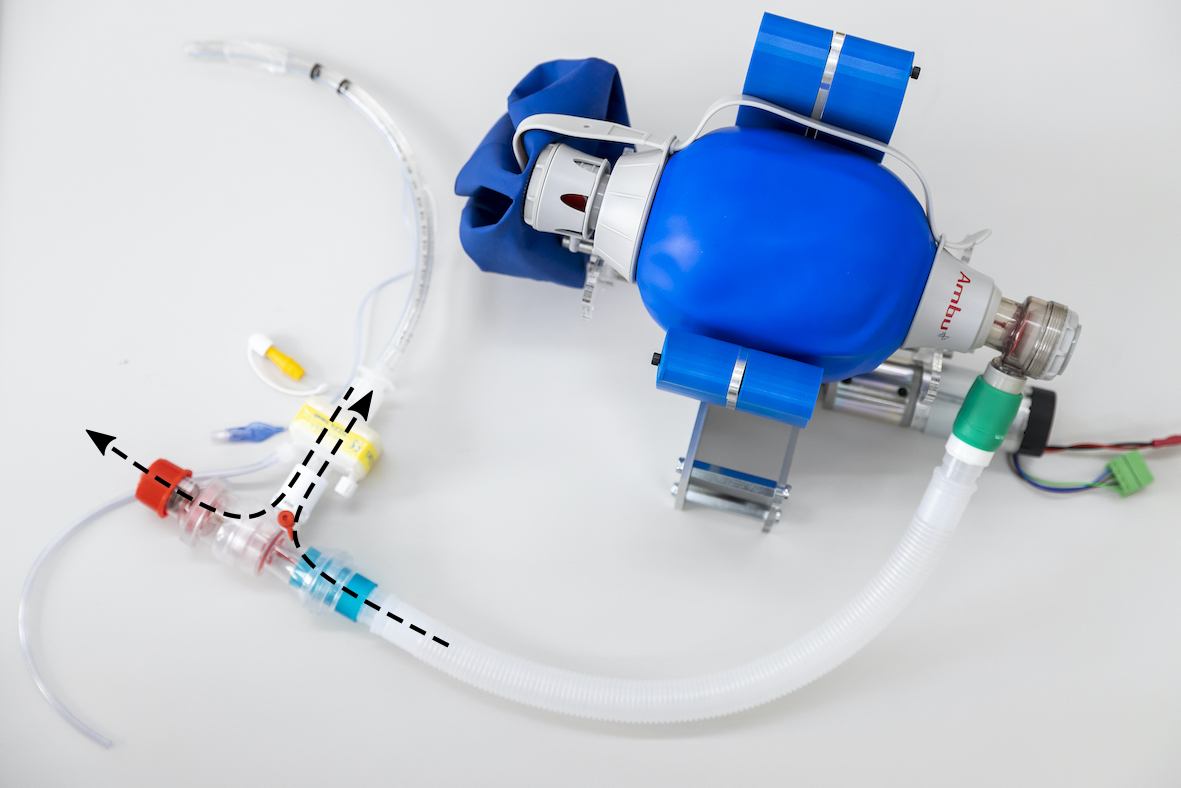}};
     \node  at (2.8,2) {\footnotesize paddels};
     \draw (2.2,1.95) -- (1.5,1.8);
     \node  at (2.8,1.3) {\footnotesize AmbuBag};
     \draw (2.05,1.25) -- (1.25,1);
     \node  at (2.7,-0.75) {\footnotesize motor};
     \draw (2.3,-0.6) -- (1.9,0);
     \node  at (-2.25,1) {\footnotesize intubation tube};
     \draw (-2,1.2) -- (-1.5,1.5);
     \node  at (-0.1,-1) {\footnotesize patient valve};
     \draw (-1, -1) -- (-1.9,-0.9);
     \node  at (-2.5,-1.7) {\footnotesize PEEP valve};
     \draw (-2.6, -1.5) -- (-2.4,-0.7);
     \node  at (-0.2,-0.1) {\footnotesize HEPA filter};
     \draw (-1.05,-0.2) -- (-1.3,-0.3);
     \node  at (-0.6,-1.7) {\footnotesize inhale};
     \node  at (-3,0) {\footnotesize exhale};
\end{tikzpicture}
      \caption{
      Functional prototype of the Breathe ventilator. 
      A resuscitator bag is squeezed by two paddles, which are actuated by an electric motor.
      The ventilator pushes air into the patient's lungs, where the breath cycle is adjusted by a range of settings. Not shown: Flow sensor (placed between the patient valve and high-efficiency particulate air (HEPA) filter) and pressure sensor (attached to HEPA filter), cf.~Fig.~\ref{fig:setup}. (Photograph: Nicola Pitaro/ETH Zurich)
      }
      \label{fig:breathe}
\end{figure}

Breathe\footnote{A short video of the Breathe ventilator and its functionalities can be found here: breathe.ethz.ch.} is a ventilator solution whose design has been adapted from \cite{MITevent}.
It is based on a resuscitator bag (Ambu Mark IV, Ambu, Ballerup, Denmark), which is squeezed by two paddles actuated by an electric motor, see Fig.~\ref{fig:breathe}.
The system is controlled with an Arduino Mega 2560.
The breathing system, i.e., the airway to the patient, consists of medical-grade tubing and  includes a bi-directional patient valve (included in the Ambu SPUR II Disposable Resuscitator, Ambu, Ballerup, Denmark), a passive PEEP valve (Ambu PEEP Valve, Ambu, Ballerup, Denmark) and a high-efficiency particulate air (HEPA) filter. The connection to the patient is realized by an intubation tube.

\subsubsection{Actuator}
The system uses a brushed $12$~V DC electric motor and a 1/212 gearbox (IG420504-SY5513, Conrad Electronic, Hirschau, Germany) with a $5.5$~A nominal current.
The motor-driver is a full H-bridge that accepts ultrasonic PWM frequencies to regulate the motor voltage. 
It also embeds an analog current-sense feedback and a digital incremental encoder that are used in the motor controller. 

\subsubsection{Sensors}
The ventilator uses two sensors: a flow meter and a pressure sensor. 
The flow meter (SFM3019, Sensirion, Staefa, Switzerland) is a 16-bit digital sensor able to measure an airflow between $-10$ and $240$~standard liter per minute. 
The sensor is temperature-calibrated and offers a $2$\% accuracy and $1$\% repeatability, which makes the measurements very reliable. 
The tidal volume is calculated by integrating the positive airflow (from the ventilator to the lung) during one breath.
The pressure sensor is a Freescale Semiconductor (MFPX5010DP, NXP Semiconductors, Eindhoven, Netherlands), which is an analog, temperature-calibrated, and pre-amplified sensor that can measure pressures in the range between $0$ to $100$~mBar. 
The modularity of the system allows the sensors (flow meter and pressure sensor) to be implemented at any position of the patient's airway. In this study, we placed the flow meter between the HEPA filter and the patient valve and connected the pressure sensor to the Luer lock port of the HEPA filter.

\begin{remark}[Sensor Placement] The placement of sensors (in particular the flow sensor) is a difficult problem on its own. We placed the flow sensor to increase the accuracy of tidal volume measurements by including potential leakage of the patient valve in the measurements. For the final placement further analysis will have to be performed in extensive device testing.
\end{remark}

\subsubsection{Alarms}
The implementation of alarms is a crucial component for low-cost ventilator solutions as outlined in the guidelines by the Medicines and Healthcare products Regulatory Agency \cite{UKguidelines}. We use the flow meter and the pressure sensor to trigger the following three alarms:
\begin{enumerate}[(a)]
\item tidal volume (as integral of positive flow) not achieved,
\item minute volume (as one minute moving average tidal volume) not achieved,
\item maximum inspiratory airway pressure exceeded, and
\item minimum inspiratory pressure not achieved,
\end{enumerate}
where excess of the pressure levels in alarm (c) also stops the current breath early to prevent overpressure, and alarm (d) also serves as a disconnection alarm.

\subsection{Control Design for the Breathe Ventilator}

The basis of the control design is a reference trajectory generation for the motor position.
The reference trajectory is a sawtooth signal composed of piecewise linear functions, whose slopes are adjusted based on the I:E and the BPM settings, as well as on the setting for the tidal volume.
The slopes, i.e., the motor velocities are given by
\begin{align*}
\dot \refpos_{\rm inhale} = \frac{\refpos_{\target} - \refpos_0}{\frac{{\rm I}}{{\rm I+E}}T_{\rm b}},\quad
\dot \refpos_{\rm exhale} = \frac{\refpos_0 - \refpos_{\target}  }{\frac{{\rm E}}{{\rm I+E}}T_{\rm b}},
\end{align*}
where $T_{\rm b}=\frac{60~{\rm s}}{{\rm BPM}}$ is the duration of one breath cycle, I and E define the I:E ratio, $\refpos_0$ is the motor position at the beginning of the inhalation phase, and $\refpos_{\target}$ is the motor position at the end of the inhalation---the set-point (sp). Throughout the paper, numeric values for the motor position are stated after gearbox reduction.
The reference trajectory is tracked using a position-velocity-current cascade of proportional-integral (PI) controllers. 
This ensures that the I:E and the BPM requirements are satisfied. 
The realized tidal volume depends on correctly choosing $\refpos_{\target}$ (with fixed $\refpos_0$).
However, for variations in patients and settings of the ventilator, $\refpos_{\target}$ has to be adapted to achieve the desired tidal volume, which will be shown with an empirical study in this paper.

In order to allow for accurate tracking of the tidal volume, we propose to use a set-point adaptation, i.e., the end-point of the position reference is altered adaptively.
Mathematically, we assume that the tidal volume of each breath, $V(k)$, results from 
\begin{align}
\label{eq:setpoint}
    V(k) = f_\theta(\refpos_{\target}(k)),
\end{align}
where $k$ is a time index that refers to the $k$th breath cycle, $\refpos_{\target}(k)$ is the motor target position of the $k$th breath cycle, and $f_\theta$ is a bounded and strictly monotonically increasing function that relates the motor target position to the tidal volume. 
We denote with $\theta$ the uncertainties arising due to patient variations, hardware variations (e.g., different resuscitator bags), or operational variations (parameter settings).
An approximation of the function $f_\theta$ based on geometric considerations is provided in Appendix~A.
Strict monotonicity of $f_\theta$ implies that the more the bag is squeezed, the higher is the achieved tidal volume, which is a reasonable assumption and was observed to be satisfied in our experiments.

The goal of the set-point adaptation algorithm is to successively adjust the target position $\refpos_{\target}(k)$ to achieve the reference tidal volume $V_{\reference}$, as quickly as possible but without overshoot. 
Hence, we want to find $\phi_{\reference}$ for which
\begin{align}
\label{eq:ref}
    V_{\reference}
    =
    f_\theta
    (\refpos_{\reference}).
\end{align}

We propose to use a set-point adaptation that is based on integrating the difference between the reference tidal volume and the achieved one, 
with
\begin{align}
\label{eq:adapt}
\refpos_{\target}(k\!+\!1)
=
\refpos_{\target}(k)
+
g_I
(
V_{\reference} - V(k)
),
\end{align}
where $g_I$ can be interpreted as a gain.
Setting $g_I=\frac{1}{\dV}$ with
\begin{align}
\label{eq:gain}
\dV
= 
\max_{\theta,\refpos_{\target}}
\left.
\frac{\partial f_\theta ( \refpos)}{\partial  \refpos}
\right|_{ \refpos = \refpos_{\target}},
\end{align}
ensures that the set point, and therefore also the tidal volume, converge to the reference value monotonically,  i.e.,
\begin{subequations}
\label{eq:monotone}
\begin{align}
 \refpos_{\reference}
\geq 
\refpos_{\target}(k\!+\!1)\geq \refpos_{\target}(k) \quad  {\rm if}\ \refpos_{\reference}\geq \refpos_{\target}(k)
\\
\refpos_{\reference}
\leq 
\refpos_{\target}(k\!+\!1) \leq \refpos_{\target}(k)\quad {\rm if}\ \refpos_{\reference} \leq \refpos_{\target}(k)
\end{align}
\end{subequations}
and asymptotically, i.e.,
\begin{align}
\label{eq:asympotic}
\refpos_{\target}(k)\rightarrow \refpos_{\reference}\ {\rm as}\ k\rightarrow \infty,
\end{align}
which is formally shown in Theorem~\ref{thm:convergence} with proof in Appendix~B.
Note that $\dV$ denotes the best Lipschitz constant or the maximum sensitivity (the steepest gradient) of the tidal volume function for all possible variations, $\theta$, and all possible positions, $\refpos$.

Empirically, we approximate the maximum sensitivity 
as
\begin{align*}
\dV
\approx 
\max_{i,j}
\frac{
f_{\theta_i}(\refpos_j + \Delta \refpos)
-
f_{\theta_i} (\refpos_j) 
}{\Delta \refpos}
\end{align*}
from a grid over a set of test cases $\{ \theta_i \}$ and motor target positions $\{ \refpos_j \}$ with step size $\Delta \refpos$.

\begin{theorem}
\label{thm:convergence}
Consider the relation between set-point and tidal volume as in \eqref{eq:setpoint} with the strictly monotonically increasing function $f_\theta$.
Using the set-point adaptation in \eqref{eq:adapt} with $g_I=\frac{1}{\dV}$ and $\dV$ as in \eqref{eq:gain}, $\refpos_{\target}$ converges monotonically and asymptotically, i.e., \eqref{eq:monotone} and \eqref{eq:asympotic} hold.
\end{theorem}

\subsection{Experimental Setup and Protocol}

\begin{figure}[t]
      \centering 
     \includegraphics[width=8cm]{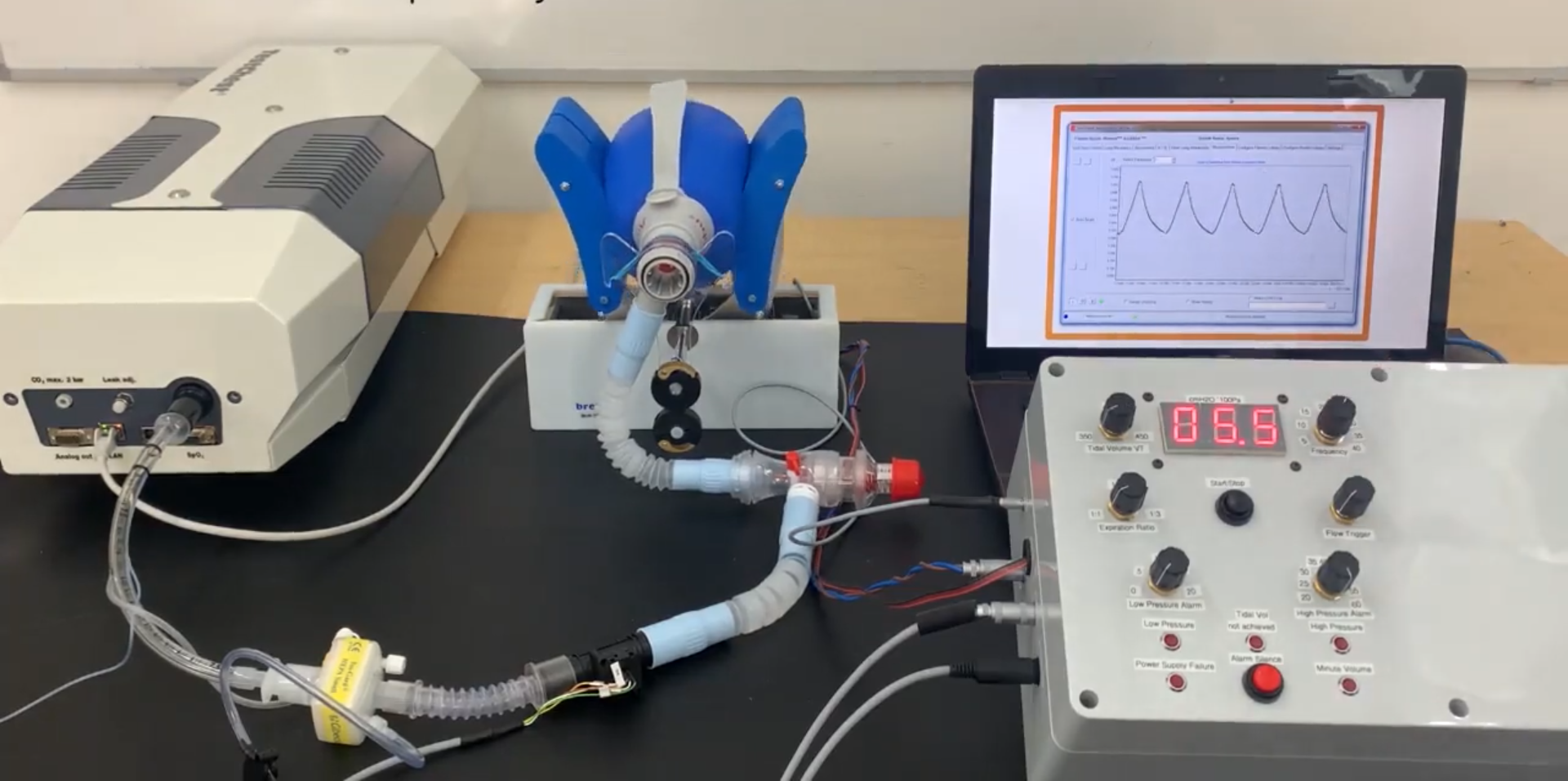}
      \caption{Experimental setup of the Breathe ventilator (center) with the TestChest (left) and the user interface (right). The ventilator was connected to the TestChest with standard medical components. The breathing system is constructed in the following order. A medical-grade air hose was used to connect the patient valve (including the positive end‐expiratory pressure (PEEP) valve) to the resuscitator bag and another, identical air hose to connect the flow sensor and the high-efficiency particulate air (HEPA) filter in line to the PEEP valve. The pressure sensor is connected to the Luer lock port of the HEPA filter. The intubation tube was attached to the HEPA filter with an adapter.
In the experiments, the inhalation path from the bag to the intubation tip had a total length of 1000~mm and a volume of approximately 195~mL. The exhalation path from the intubation tip to the PEEP valve had a length of 765~mm and a volume of approximately 120~mL.}
      \label{fig:setup}
\end{figure}

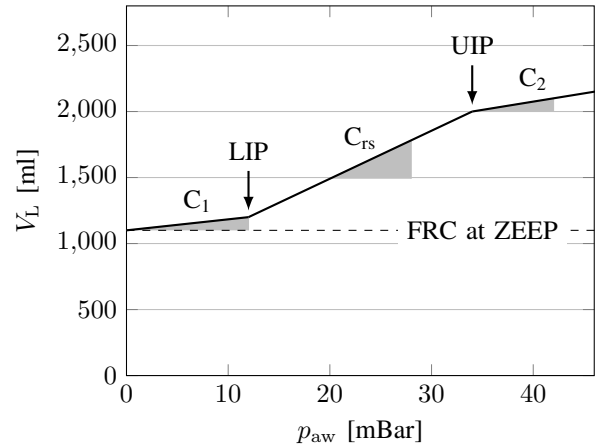
\begin{figure}[t]
\centering 
\pgfplotsset{compat=newest}

\begin{tikzpicture}[every plot/.append style={ thick}]
\begin{axis}[ 
   xmin=0, xmax=46,
ymin=0, ymax=2800, 
ytick = {0,500,1000,1500,2000,2500,3000},
xlabel = {$p_{\rm aw}$ [mBar]},
ylabel = {$V_{\rm L}$ [ml]}, 
 ylabel near ticks,
    legend pos=outer north east, 
    legend style={at={(1,0)},anchor=south east,},
    ymajorgrids=true,
    yminorgrids=true,
    legend columns=5,
    width=7.8cm,
    height=6.5cm,
  ]
] 

\node[] at (7,1320) {C\textsubscript{1}};
\draw[draw=lightgray,fill=lightgray] (0,1100) -- (12,1200) -- (12,1100);
\node[] at (40,2250) {C\textsubscript{2}};
\draw[draw=lightgray,fill=lightgray] (34,2000) -- (42,2100) -- (42,2000);
\node[] at (23,1800) {C\textsubscript{rs}};
\draw[draw=lightgray,fill=lightgray] (20,1200+291) -- (28,1200+582) -- (28,1200+291);

\draw[black,dashed] (0,1100) -- (50,1100) {};
\draw[black,thick] (0,1100) -- (12,1200) {};
\draw[black,thick] (34,2000) -- (12,1200) {};
\draw[black,thick] (34,2000) -- (50,2200) {};

\node[fill=white] at (35,1100) {FRC at ZEEP};
\node[] at (12,1700) {LIP};
\draw[thick, ->,>=latex] (12,1550) -- (12, 1250);
\node[] at (34,2500) {UIP};
\draw[thick, ->,>=latex] (34,2350) -- (34, 2050);

\end{axis}
\end{tikzpicture}
      \caption{Illustration of the basic lung mechanics, relating the lung volume ($V_{\rm L}$) to the airway pressure ($p_{\rm aw}$), see \cite{Neosim2020Model}. 
      C\textsubscript{1} is the respiratory compliance below the lower inflection point (LIP),
      C\textsubscript{2} is the respiratory compliance above the upper inflection point (UIP), and
      C\textsubscript{rs} is the respiratory compliance between the LIP and the UIP. 
      FRC at ZEEP denotes the functional residual capacity at zero end-expiratory pressure.
      }
      \label{fig:lungmech}
\end{figure}

For the experimental testing, the ventilator was connected to a mechanical lung, the TestChest via intubation with an inflated cuff, see Fig.~\ref{fig:setup}. 
The TestChest is specially designed for ventilator training and has an underlying lung model integrated. The basic lung mechanics are defined by the pressure-volume curve for static conditions in Fig.~\ref{fig:lungmech} and the airway resistance, $R_{\rm aw}$. The respiratory compliance, C\textsubscript{rs}, is determined by the chestwall compliance, C\textsubscript{w}, and the lung compliance, C\textsubscript{L}, as ${\rm 1/C_{rs} = 1/C_{w} + 1/C_{L}}$.

Table~\ref{tb:chest_settings} shows two sets of TestChest settings, one set modeling a healthy person and one set modeling a patient suffering from COVID-19. All physiological values were provided by \cite{Neosim2020} and were implemented in the physiological lung model of the TestChest. The basic reference values of lung characteristics of simulated patients were summarized by \cite{Arnal2018parameters}. The assumption that late-stage COVID-19 type H patients match the severe acute respiratory distress syndrome (ARDS) criteria was emphasized by \cite{Gattinoni2020pneumonia}.

In order to show the required set-point adaptation for individual lung characteristics, the following base settings were chosen for the experiments. The tidal volume was measured as the time integral of the flow meter on the disconnected ventilator and on both patient models of the TestChest (healthy and ARDS) with a PEEP of 5~mBar. Additionally, the PEEP valve was set to 10~mBar for the ARDS patient to highlight that different ventilator settings require an adaptation of the set-point.

\renewcommand{\arraystretch}{1.05}

\begin{table}[!t]
    \caption{}
  \vspace{-0.35cm}
  Lung Mechanics setting in the TestChest to simulate a healthy subject and a patient suffering from acute respiratory distress syndrome (ARDS) due to COVID-19.
  \center
    \begin{tabular}{cccc} \toprule
         Parameter & [Unit] & Healthy & ARDS  
         \\ \midrule
         C\textsubscript{w} & [mL/mBar] & 200 & 93 \\
         $R_{\rm aw}$ & [mBar/(L/s)]    & Rp5  & Rp5 \\
         FRC at ZEEP &  [mL] & 2000 & 1102 \\
         LIP & [mBar] & 5 & 12 \\
         UIP & [mBar] & 35 & 35 \\
         C\textsubscript{1} & [mL/mBar]& 25 & 8 \\
         C\textsubscript{rs} & [mL/mBar] & 60 & 35 \\
         C\textsubscript{2} & [mL/mBar]& 20 & 8 \\
         \bottomrule
    \end{tabular}
    \label{tb:chest_settings}
\end{table}

\section{Results}

\subsection{Parameter Study with fixed Set-Point}
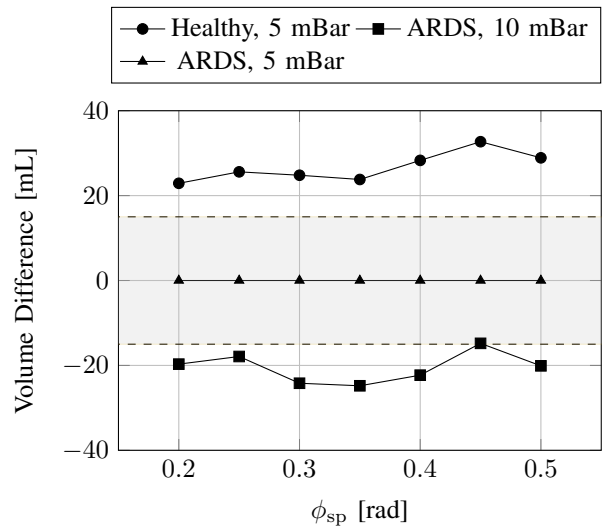
\begin{figure}[t]
    \centering 
    \begin{tikzpicture}
	\begin{axis}[
		height=6.1cm,
		width=8cm,
		grid=major,
		xlabel = $\refpos_{\target}$ {[rad]},
		ylabel = Volume Difference {[mL]},
		xmax = 0.55,
		xmin = 0.15,
		ymin = -40,
		ymax = 40,
		legend style={at={(1,1.08)},anchor=south east,},
		legend columns=2
	]

	\addplot[color=black,
                solid,
                mark=*,] coordinates {
		(0.2,22.9)
		(0.25,25.6)
		(0.3,24.8)
		(0.35,23.8)
		(0.4,28.3)
		(0.45,32.7)
		(0.5,28.9)
	};
	
	\addplot[color=black,
                solid,
                mark=square*] coordinates {
		(0.2,-19.7)
		(0.25,-17.9)
		(0.3,-24.2)
		(0.35,-24.8)
		(0.4,-22.3)
		(0.45,-14.8)
		(0.5,-20.1)
	};

	\addplot[color=black,
                solid,
                mark=triangle*,] coordinates {
		(0.2,0)
		(0.25,0)
		(0.3,0)
		(0.35,0)
		(0.4,0)
		(0.45,0)
		(0.5,0)
	};
	
	\addplot[dashed, black] coordinates {
		(0,-15)
		(1,-15)
	};
	
		\addplot[dashed, black] coordinates {
		(0,15)
		(1,15)
	};
		
		\addplot[patch, patch type = rectangle, lightgray, opacity = 0.2] table {
		x y
		0 15
		1 15
		1 -15
		0 -15
	};

	\addlegendentry{Healthy, 5 mBar}
	\addlegendentry{ARDS, 10 mBar}
	\addlegendentry{ARDS, 5 mBar}
	\end{axis}
\end{tikzpicture}
    \caption{Tidal volume difference to acute respiratory distress syndrome (ARDS) patient with 5~mBar over the target motor position set-points for variation in positive end-expiratory pressure (PEEP) and patient specifications (20~breaths per minute (BPM), Inspiratory:Expiratory ratio (I:E) = 1:2). 
    The gray-shaded area shows that all scenarios considered in the study are separated by more than 15~mL. The scenario in which the ventilator was disconnected from the TestChest is not displayed as the volume differences are greater than 40~mL for all target motor positions.}
    \label{fig:tidel_diff}
\end{figure}

In a parameter study, we investigated the implications of different settings and different patients on the tidal volume for a fixed set-point, $\phi_{\target}$.
The resulting tidal volumes for the various test scenarios of different patient settings, PEEP settings, and set-points are reported in Table~\ref{tb:param_study}.
The parameter study showed that the tidal volume varied greatly for a fixed set-point.
For small set-points, the tidal volume varied more than 30\% with different patient characteristics and PEEP settings, 
e.g., 
$101~{\rm mL}/58~{\rm mL}\approx 174\%$ for $\phi_{\target}=0.2~{\rm rad}$ or
$161~{\rm mL}/118~{\rm mL}\approx 136\%$ for $\phi_{\target}=0.25~{\rm rad}$.
For higher set-points, the tidal volume varied more than 10\%,
e.g., 
$503~{\rm mL}/453~{\rm mL}\approx 111\%$ for $\phi_{\target}=0.5~{\rm rad}$ or
$444~{\rm mL}/397~{\rm mL}\approx 111\%$ for $\phi_{\target}=0.45~{\rm rad}$.
Fig.~\ref{fig:tidel_diff} illustrates tidal volume changes for the test scenarios as in Table~\ref{tb:param_study} by displaying the difference to the ARDS case with $5$~mBar PEEP. 
It highlights that the tidal volumes for various patients or settings with a fixed set-point 
vary by more than $\pm 15$~mL.
This analysis indicates that there is no single fixed set-point to achieve a given tidal volume for the various setups considered in the study.

\setlength{\tabcolsep}{7pt}
\begin{table}[t]
  \caption{}
  \vspace{-0.35cm}
  Tidal volumes for varying set-points, different configurations (disconnected from the TestChest, healthy TestChest patient, acute respiratory distress syndrome (ARDS) TestChest patient), and 
  PEEP valve settings, as average values over 10 breath cycles (20 breaths per minute (BPM), Inspiratory:Expiratory ratio (I:E) = 1:2). 
  \center
  \label{tb:results}
  \begin{tabular}{c|cccc} \toprule
  &  \multicolumn{4}{c}{Tidal Volume [mL]} \\
   	& & Healthy   & ARDS  & ARDS  \\  
 $\refpos_{\target}$ {[rad]} & discon. & 5\,mBar & 5\,mBar & 10\,mBar \\ \midrule 
0.20 & 197 & 101 &  78 & 58 \\ 
0.25 & 260 & 161 &  136 & 118 \\ 
0.30 &  340 & 225 &  200 & 176 \\ 
0.35 & 428 & 292 &  268 & 243 \\ 
0.40 & 488 & 365 &  338 & 315 \\ 
0.45 & 506 & 444 &  412 & 397 \\ 
0.50 & 517 & 503 &  474 & 453 \\
         \bottomrule
  \end{tabular}
\label{tb:param_study}
\end{table}

\subsection{Set-Point Adaptation}

In the second study, we investigated the implications of the set-point adaptation on the achieved tidal volume. 
We estimated $\dV$ in \eqref{eq:gain} with the results in Table~\ref{tb:param_study}, resulting in $g_I = 5.6 \cdot 10^{-4}$~[rad/mL], which we rounded down to $5 \cdot 10^{-4}$ for additional safety.
Table~\ref{tb:results_2} shows the tracking performance with the set-point adaptation, i.e., $\phi_{\target}$ was adjusted according to \eqref{eq:adapt}. 
For all combinations of tidal volumes and BPM, the tidal volume was tracked very closely with low mean error and root mean squared error (RMS) of approximately 1~mL and a maximum deviation of the tidal volumes from the target of less than 4~mL (absolute value).
In order to better judge the maximum and the RMS values, we conducted one experiment without adaptation using representative settings  
(ARDS patient, 100 breath cycles, $V_{\reference}=400$~mL, 20~BPM), which resulted in a maximum deviation of 2.24~mL and RMS of 0.71~mL. We therefore observed repeatability of tidal volume measurements within 0.2\%.  

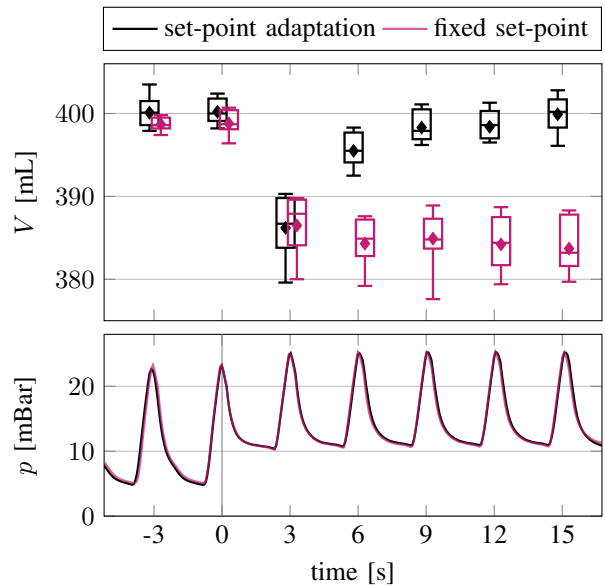
\begin{figure} 
\centering
\pgfplotsset{compat=newest}
\usepgfplotslibrary{statistics}

\newcommand{\figheight}{5cm}
\newcommand{\figwidth}{8.2cm}
\newcommand{\evpo}{2}
 
\definecolor{mst1}{rgb}{0.7629,    0.1073,    0.4411}
\definecolor{mst2}{rgb}{0.9665,    0.6340,   0.1526}
\definecolor{mst3}{rgb}{   0.7629,    0.1073,    0.4411}

\begin{tabular}[t]{rrr}
\begin{tikzpicture} 
\begin{axis}[
   xmin=-6, xmax=15.95,
ymin=375, ymax=406, 
xtick = {-3.8,-0.8,2.2,5.2,8.2,11.2,14.2},
xticklabels = {,,,,,,,,,,,,}, 
ylabel = {$V$ [mL]},
    boxplot/draw direction=y,
 ylabel near ticks,
    legend pos=outer north east,  
    legend style={at={(1,1.05)},anchor=south east,}, 
    ymajorgrids=true,
    legend columns=5, 
    width=\figwidth,
    height=\figheight,
  ]
] 

\addplot[line width=0.3mm ,color=black] table[x index=0,y index=1,each nth point={\evpo}]  {data_p_Ad_mean.txt};

\addplot[line width=0.3mm ,color=mst1,opacity=0.7] table[x index=0,y index=1,each nth point={\evpo}]  {data_p_noAd_mean.txt}; 

\legend{set-point adaptation,fixed set-point}

          \addplot+[mark = *, line width=0.3mm ,solid,mark options = {mst1},
    boxplot prepared={
    draw position=-4,
      lower whisker=397.9,
      lower quartile=398.6,
      median=400.1,
      average=400.1,
      upper quartile=401.5,
      upper whisker=403.5
    }, color = black
    ] coordinates{
    };    
             \addplot+[mark = *, line width=0.3mm ,solid,mark options = {mst1},
    boxplot prepared={
    draw position=-1,
      lower whisker=398.2,
      lower quartile=399.1,
      median=400,
      average=400.2,
      upper quartile=401.8,
      upper whisker=402.4
    }, color = black
    ] coordinates{
    };    
     \addplot+[mark = *, line width=0.3mm ,solid,mark options = {mst1},
    boxplot prepared={
    draw position=2,
      lower whisker=379.6,
      lower quartile=383.8,
      median=386.7,
      average=386.2,
      upper quartile=389.8,
      upper whisker=390.3
    }, color = black
    ] coordinates{
    };     
     \addplot+[mark = *, line width=0.3mm ,solid,mark options = {mst1},
    boxplot prepared={
    draw position=5,
      lower whisker=392.5,
      lower quartile=394.1,
      median=395.5,
      average=395.5,
      upper quartile=397.7,
      upper whisker=398.3
    }, color = black
    ] coordinates{
    };     
     \addplot+[mark = *, line width=0.3mm ,solid,mark options = {mst1},
    boxplot prepared={
    draw position=8,
      lower whisker=396.2,
      lower quartile=396.9,
      median=397.9,
      average=398.3,
      upper quartile=400.5,
      upper whisker=401.1
    }, color = black
    ] coordinates{
    };    
     \addplot+[mark = *, line width=0.3mm ,solid,mark options = {mst1},
    boxplot prepared={
    draw position=11,
      lower whisker=396.5,
      lower quartile=397,
      median=398.6,
      average=398.4,
      upper quartile=400.3,
      upper whisker=401.3
    }, color = black
    ] coordinates{
    };    
     \addplot+[mark = *, line width=0.3mm ,solid,mark options = {mst1},
    boxplot prepared={
    draw position=14,
      lower whisker=396.1,
      lower quartile=398.3,
      median=400.2,
      average=399.9,
      upper quartile=401.7,
      upper whisker=402.8
    }, color = black
    ] coordinates{
    };

    \addplot+[mark = *, line width=0.3mm ,solid,mark options = {mst1},
    boxplot prepared={
    draw position=-3.5,
      lower whisker=397.4,
      lower quartile=398.2,
      median=398.6,
      average=398.7,
      upper quartile=399.5,
      upper whisker=399.8
    }, color = mst1
    ] coordinates{
    };   
    \addplot+[mark = *, line width=0.3mm ,solid,mark options = {mst1},
    boxplot prepared={
    draw position=-0.5,
      lower whisker=396.4,
      lower quartile=398.1,
      median=398.7,
      average=398.8,
      upper quartile=400.4,
      upper whisker=400.7
    }, color = mst1
    ] coordinates{
    };   
    \addplot+[mark = *, line width=0.3mm ,solid,mark options = {mst1},
    boxplot prepared={
    draw position=2.5,
      lower whisker=380,
      lower quartile=384.1,
      median=387.9,
      average=386.5,
      upper quartile=389.6,
      upper whisker=389.8
    }, color = mst1
    ] coordinates{
    };   
    \addplot+[mark = *, line width=0.3mm ,solid,mark options = {mst1},
    boxplot prepared={
    draw position=5.5,
      lower whisker=379.2,
      lower quartile=382.8,
      median=384.9,
      average=384.3,
      upper quartile=387.2,
      upper whisker=387.6
    }, color = mst1
    ] coordinates{
    };   
    \addplot+[mark = *, line width=0.3mm ,solid,mark options = {mst1},
    boxplot prepared={
    draw position=8.5,
      lower whisker=377.6,
      lower quartile=383.7,
      median=384.8,
      average=384.9,
      upper quartile=387.3,
      upper whisker=388.9
    }, color = mst1
    ] coordinates{
    };   
    \addplot+[mark = *, line width=0.3mm ,solid,mark options = {mst1},
    boxplot prepared={
    draw position=11.5,
      lower whisker=379.4,
      lower quartile=381.7,
      median=384.4,
      average=384.2,
      upper quartile=387.5,
      upper whisker=388.7
    }, color = mst1
    ] coordinates{
    };   
    \addplot+[mark = *, line width=0.3mm ,solid,mark options = {mst1},
    boxplot prepared={
    draw position=14.5,
      lower whisker=379.7,
      lower quartile=381.6,
      median=383.2,
      average=383.7,
      upper quartile=387.8,
      upper whisker=388.3
    }, color = mst1
    ] coordinates{
    };

\end{axis}

\end{tikzpicture}
\\[-0.2cm]

\begin{tikzpicture} 
\begin{axis}[ 
   xmin=-7.2, xmax=14.75,
ymin=0, ymax=28, 
xtick = {-3,0,3,6,9,12,15},
xtick = {-5,-2,1,4,7,10,13},
xticklabels = {-3,0,3,6,9,12,15},
ytick={0,10,20},
yticklabels={$0$,$10$,$\phantom{1}{20}$},
xlabel = {time [s]},
ylabel = {$p$ [mBar]},  
    boxplot/draw direction=y,
 ylabel near ticks,
    legend pos=outer north east,  
    legend style={at={(1,2.5)},anchor=south east,}, 
    ymajorgrids=true, 
    legend columns=5, 
    width=\figwidth,
    height=0.8*\figheight,
  ]
]

\draw[gray] (-2,0) -- (-2,30) {}; 
\addplot[line width=0.3mm ,color=black] table[x index=0,y index=1,each nth point={\evpo}]  {data_p_Ad_mean.txt};
\addplot[line width=0.3mm ,color=mst1,opacity=0.7] table[x index=0,y index=1,each nth point={\evpo}]  {data_p_noAd_mean.txt};
 
\end{axis}

\end{tikzpicture}
\end{tabular}
 
\caption{Effects of positive End-Expiratory Pressure (PEEP) valve change from 5 to 10 mBar on tidal volume and pressure with and without set-point adaptation, with 10 repetitions of the experiment (acute respiratory distress syndrome (ARDS) patient, tidal volume setpoint $V_{\reference}=400$~mL, 20~breaths per minute (BPM), Inspiratory:Expiratory ratio (I:E)~=~1:2).  The change occurred at time 0~s.
Top: Statistics of the measured tidal volume, $V$, computed as the time integral of the positive flow for each breath and represented as boxplots.
The diamond symbol and the vertical line represent the mean and the median, respectively; the box edges represent the 75th and the 25th percentiles, and the whiskers represent the maximum as well as the minimum value. Note that the high variance during the first breath after the change is likely due to inaccuracies during the manual adjustments of the PEEP valve.
Bottom: Mean over all pressure trajectories as an indication of PEEP. 
}
\label{fig:transient}
\end{figure}

To illustrate the transient behavior of the adaptation mechanism, we changed the operating condition in the ARDS patient setting through a manual adjustment of the PEEP valve from 5 to 10~mBar. Fig.~\ref{fig:transient} displays the resulting tidal volumes of 5 breath cycles following the change with and without set-point adaptation. It shows that the adaptation required roughly two to three breath cycles to achieve the desired tidal volume.

\begin{remark}
Results for further BPM and I:E settings were qualitatively similar, but are not reported here.
\end{remark}

\begin{table}[t] 
  \caption{} 
  \label{tb:results_2}
  \vspace{-0.35cm}
  Tidal volume tracking error for different target tidal volumes and breaths per minute (BPM), averaged over 100 breath cycles for each setting (Acute respiratory distress syndrome (ARDS) patient, 5~mBar, Inspiratory:Expiratory ratio (I:E) = 1:2). 
  \center 
\begin{tabular}{cc|ccc} \toprule
& & \multicolumn{3}{c}{ Tidal volume tracking error  
}\\
& & \multicolumn{3}{c}{ $\Delta V = V_{\text{ref}} - V(k)$ [mL]}\\
$V_{\text{ref}}$  {[mL]} & BPM & \hspace{0.1cm} Mean \hspace{0.1cm}  &  RMS  &   Max. (abs.) 
\\ \midrule 
350 & 10 & 0.01 & 1.12 & 3.26 \\
    & 20 & -0.01 & 0.78 & 1.90\\
    & 30 & 0.03 & 0.74 & 1.80 \\  \midrule 
400 & 10 & -0.04 & 1.05 & 2.83\\
    & 20 & 0.06 & 0.83 & 3.69\\
    & 30 & -0.01 & 0.84 & 2.85 \\  \midrule 
450 & 10 & -0.24& 1.14 & 3.58\\
    & 20 & -0.0 & 0.90 & 2.84\\
    & 30 & 0.03 & 0.77 & 2.51\\    
         \bottomrule
  \end{tabular}
\end{table}

\section{Discussion}

Low-cost mechanical ventilator solutions that use paddles to squeeze a  resuscitator bag (AmbuBag) offer sufficient flexibility to provide a wide range of operating settings. 
However, the requirements for their medical usage render a ventilator design without a flow meter and a pressure sensor difficult, e.g., for triggering alarms and for the controller. 

In this study, we used a high-fidelity experimental testbed to show that variations in patient pathophysiology, in the hardware, or in operational parameter settings cause significant differences in the tidal volume if the controller is not adjusted accordingly.
These differences are likely due to varying compression and leakage losses in the breathing system, which will tend to be particularly pronounced and irregular in a low-cost and modular solution as we consider here.
In particular, the parameter study showed that no two test scenarios with different settings or patients were within a range of $\pm 15$~mL, with maximum deviations of $50$~mL.
The results indicate that the feeding back of sensor measurements is necessary to adjust the controller in order to accurately track the desired tidal volume for different patients and operating conditions.

The proposed set-point adaptation was shown to cope with different settings and patients by automatically adapting the controller to achieve the desired tidal volume.
The controller using the set-point adaptation was able to stay within a tolerated range of $\pm 10$~mL for all breath cycles test scenarios. 
The variations over the experiments (RMS and maximum tracking errors) were found to be on a similar scale both with and without set-point adaptation, which indicates that the set-point adaptation is not sensitive to noise.
Furthermore, the analysis of transients due to changes in the operating conditions indicate that the set-point adaptation is sufficiently fast to react within two to three breath cycles.

In principle, it would be possible to store different set-points for different operating conditions, different resuscitator bags, and different patients. 
However, providing a set-point for all possible combinations and to select them for each patient separately would be a great overhead for the engineer of the ventilator and the doctors/nurses.

\section{Conclusion}

This paper experimentally investigated the implications of different patient pathophysiologies and different operating conditions on low-cost ventilator solutions that compress a bag to pump air into a patient's lungs.
A parameter study highlighted the need for flow measurements and an adaptive solution for a controller to achieve the high accuracy for different patients and different operating conditions, in particular the tidal volume. 
Furthermore, this paper proposed a set-point adaptation algorithm to automatically adjust the controller to different scenarios. 
The set-point adaptation algorithm was shown to achieve the desired set points within a tolerance of less than $10$~mL.
The results indicate that such an adaptive scheme is able to cope with hardware and patient variations and should be considered when designing low-cost ventilators based on a bag-squeezing mechanism.

\section*{Acknowledgment}
 
We gratefully acknowledge Sara Mettler and Dario Fenner from the Product Development Group Zurich, ETH Zurich  for their technical support; Adrian Marty and the team from the Simulation Center at the University Hospital Zurich for providing medical knowledge and the testing platform "TestChest"; as well as Martin Meier from the Z\"urcher Hochschule der K\"unste.

\addtolength{\textheight}{-2cm}

\section*{Appendix A: Geometric Model for Resuscitator Bag}

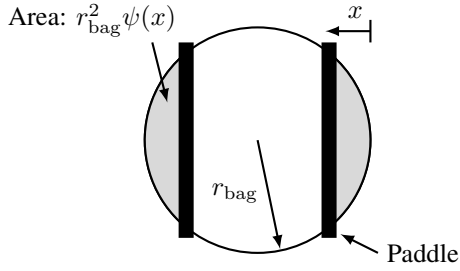
\begin{figure}[!t]
\centering
\newcommand{\radiusambu}{1.5}
\begin{tikzpicture}
\draw[thick] (0,0) circle (\radiusambu);
\draw[ thick,fill=gray!30] 
(230:\radiusambu) arc(230:130:\radiusambu) -- cycle;
\draw[ thick,fill=gray!30] 
(50:\radiusambu)--(-50:\radiusambu) arc (-50:50:\radiusambu)--cycle;

\node[] at (-2.2,1.6) {Area: $r_{\ambu}^2 \psi(x)$};
\draw[thick, ->,>=latex] (-1.4,1.3) -- (-1.2,0.5);
\draw[thick] (1.5,1.3) -- (1.5,1.6);
\draw[thick, ->,>=latex] (1.5,1.45) -- (0.9,1.45);
\node[] at (1.3,1.7) {$x$};

\draw[thick, ->,>=latex] (0,0) -- (0.3,-1.45);
\node[] at (-0.3,-0.7) {$r_{\ambu}$};

\draw[line width=2mm,] (0.95,-1.3) -- (0.95,1.3);
\draw[line width=2mm,] (-0.95,-1.3) -- (-0.95,1.3);
\node[] at (2.2,-1.5) {Paddle};
\draw[thick, ->,>=latex] (1.6,-1.5) -- (1.1,-1.25);

\end{tikzpicture}
      \caption{Illustration of geometric resuscitator bag model. The bag is modeled as cylindrical object, which is squeezed by two paddles. The resulting volume is the volume of a cylinder minus the volume displaced by the two paddles.}
      \label{fig:geometry}
\end{figure}

We model the bag as a cylindrical object, 
with radius $r_{\ambu}$ and length $l_{\ambu}$, and the lung as a constant volume, $V_{\lung}$.
Let $x$ be the horizontal distance of one paddle to the outside of the cylindrical object, see Fig.~\ref{fig:geometry} for an illustration of the geometry.
Further, let $\bar x = \frac{r_{\ambu}-x}{r_{\ambu}}$.
Then, the volume of the cylindrical object with paddle position $x$ is
\begin{align*}
V_{\ambu}(x)
&=
    l_{\ambu} r_{\ambu}^2 
    \left(
    \pi - 2 \psi(x)
    \right)
    \\
\psi(x)
&= \arccos \bar x -  \bar x\sqrt{1-\bar x^2}.
\end{align*}
Approximating the horizontal position as $x = l_{\pedal}\refpos$ with the length of the paddle $l_{\pedal}$ (distance of outside of bag to the paddle's rotation axis) and  the motor position $\refpos$, the tidal volume is given by
\begin{align*}
f_\theta(\refpos_{\target}) 
=
V_{\ambu}(l_{\pedal}\refpos_0)
-
V_{\ambu}(l_{\pedal}\refpos_{\target}).
\end{align*}

\begin{remark}
This geometric model can also be used to approximately predict the pressure and the flow rate, as well as the torque transmitted from the paddles to the motor.
The pressure, the flow rate, and the torque of one paddle are given by
\begin{align*}
    p  
    &=
    p_\infty 
    \frac{
    V_{\lung} + V_{\ambu}(0)
    }{V_{\lung} + V_{\ambu}(l_{\pedal}\refpos)}
    \\
    \dot V_{\ambu} 
    &=
    4l_{\ambu} r_{\ambu}^2 \dot{\bar{x}}
    \sqrt{1-\bar x^2}
    \\
    \tau
    &= l_{\pedal} 
    A_{\pedal}p  
\end{align*}
with the surface of the paddle computed as $A_{\pedal} =2 l_{\ambu} r_{\ambu} \sqrt{1-\bar x^2}$.
\end{remark}

\section*{Appendix B: Proof of Theorem~\ref{thm:convergence}}

\begin{proof}
As $f_\theta$ is a monotonically increasing function and $\dV \geq \frac{\partial f_\theta ( \refpos)}{\partial  \refpos}$ for any $\refpos$ and $\theta$, 
\begin{align*}
V_1 + \dV (\refpos_2-\refpos_1)
\geq V_2\quad 
{\rm if}\ \refpos_2\geq \refpos_1
\\
V_1 + \dV (\refpos_2-\refpos_1)
\leq V_2\quad 
{\rm if}\ \refpos_2 \leq \refpos_1
\end{align*}
for any two $V_2=f_\theta(\refpos_2)$ and $V_1=f_\theta(\refpos_1)$.
In particular,
\begin{align*}
V(k) + \dV (\refpos_{\reference}-\refpos_{\target}(k))
\geq V_{\reference}\quad {\rm if}\ \refpos_{\reference}\geq \refpos_{\target}(k)\
\\
V(k) + \dV (\refpos_{\reference}-\refpos_{\target}(k))
\leq V_{\reference}\quad {\rm if}\ \refpos_{\reference} \leq \refpos_{\target}(k),
\end{align*}
which is equivalently written as
\begin{align*}
\refpos_{\reference}
\geq 
\refpos_{\target}(k)+
\frac{1}{ \dV}
(V_{\reference} - V(k) )\quad  {\rm if}\ 
\refpos_{\reference}\geq \refpos_{\target}(k) 
\\
\refpos_{\reference}
\leq 
\refpos_{\target}(k)+
\frac{1}{ \dV}
(V_{\reference} - V(k) )\quad {\rm if}\ 
\refpos_{\reference} \leq \refpos_{\target}(k) 
\end{align*}
as $\dV>0$. 
Finally, choosing $\refpos_{\target}(k\!+\!1)$ as in \eqref{eq:adapt} with $g_I=\frac{1}{\dV}$ and $\dV$ as in \eqref{eq:gain}, we can conclude that \eqref{eq:monotone} holds, with $\refpos_{\target}(k\!+\!1)=\refpos_{\target}(k) \iff \refpos_{\target}(k)=\refpos_{\reference}$, which proves \eqref{eq:asympotic}. 
\end{proof}

\bibliographystyle{IEEEtran}
\bibliography{main.bbl}

\end{document}